\documentclass[english]{article}
\usepackage[dvips]{graphicx}
\usepackage{babel}
\usepackage{inputenc}
\date{}
\title{\Large{\bf{Rare Particle Searches with the high altitude SLIM
      experiment}}} 
\author{
S. Balestra $^1$, S. Cecchini$^{1,2}$,
F. Fabbri $^1$, G. Giacomelli$^1$, R. Giacomelli$^1$,\\
 M. Giorgini$^1$, A. Kumar$^{1,3}$, S. Manzoor$^{1,4}$, J. McDonald
 $^5$, A. Margiotta$^1$,\\ 
E. Medinaceli $^1$, J. Nogales $^6$, 
L. Patrizii $^1$, J. Pinfold $^5$, V. Popa $^{1,7}$,\\ I. Qureshi $^4$,
O. Saavedra $^8$, G. Sher $^4$, M. Shahzad $^4$, \\
M. Spurio $^1$, R. Ticona $^6$, V. Togo $^1$, A. Velarde $^6$,
A. Zanini $^8$.\\ 
\\
\normalsize{
(1) Dip. Fisica dell'Universit\'a di Bologna and INFN, 40127}
\normalsize{Bologna, Italy} \\  
\normalsize{(2) INAF/IASF Sez. Bologna, 40129 Bologna, Italy} \\
\normalsize{(3) Physics Dept., Sant Longowal Institute of Eng. \& Tech.,}
\normalsize{Longowal, 148 106 India}\\
\normalsize{(4) PRD, PINSTECH, P.O. Nilore, Islamabad, Pakistan}\\
\normalsize{(5) Centre for Subatomic Research, Univ. of Alberta, Edmonton,}
\normalsize{Alberta T6G 2N4, Canada}\\
\normalsize{(6) Laboratorio de Fisica Cosmica de Chacaltaya, UMSA, La Paz,}
\normalsize{Bolivia}\\
\normalsize{(7) Institute for Space Sciences, R-77125, Bucharest-M\u{a}gurele,
Romania} \\
\normalsize{(8) Dip. Fisica Sperimentale e Generale, Universit\'a di}
\normalsize{Torino and INFN, 10125 Torino, Italy}}

\begin{document}
\maketitle

\begin{quote}

\large{
Paper presented by G. Giacomelli for the SLIM collaboration at the
\emph{European Conference on High Energy Physics}, July 21 - 27 2005,
Lisboa, Portugal.}
\end{quote}
\normalsize{
\abstract{
The search for rare particles in the cosmic radiation remains one of the
main aims of non-accelerator particle astrophysics. Experiments at
high altitude allow lower mass thresholds with respect to detectors at
sea level or underground. The SLIM experiment is a large array of
nuclear track detectors located at the Chacaltaya High Altitude
Laboratory (5290 m a.s.l.). The preliminary results from the analysis
of the first 236 m$^2$ exposed for more than 3.6 y are here
reported. The detector is sensitive to Intermediate Mass Magnetic
Monopoles, 10$^5~<~m_M~<~10^{12}$ GeV, and to SQM nuggets and
Q-balls, which are possible Dark Matter candidates.}  
} 
\section{Introduction}
\large{ 
Grand Unified Theories (GUT) of the strong and 
electroweak interactions at the scale M$_G$ $\sim$
10$^{14}$ GeV predict the existence of magnetic monopoles (MMs), 
produced  
in the early Universe at the end of the GUT epoch, with 
very large masses, M$_{M} >$ 10$^{16}$ GeV. GUT poles in the cosmic 
radiation should be 
characterized by low velocity and 
relatively large energy losses \cite{MMs}. At present the MACRO
experiment has set the best limit on GUT MMs for 4 10$^{-5}$
$<\beta=v$/c$< $ 0.5 \cite{MACRO}.

Intermediate Mass Monopoles (IMMs)
[10$^{5}$ $\div $ 10$^{12}$ GeV] with magnetic charge g = 2 g$_{D}$
could also be present in the cosmic radiation; they may have been
produced in later phase transitions in the early Universe
\cite{IMMs}. The recent interest in IMMs is also
connected with the possibility that they could yield the highest
energy cosmic rays \cite{UHECR}. IMMs may have
relativistic velocities since they could be accelerated to high
velocities in one coherent domain of the galactic magnetic field. In
this case one would have to look for downgoing fast ($\beta>$0.1)
heavily ionizing MMs.
 
Relatively low mass classical Dirac monopoles
are being searched for mainly at high energy accelerators
\cite{bertani}. 

Besides MMs, other massive particles have been hypothesized to exist 
in the cosmic radiation and possibly to be components of the galactic
cold dark matter: nuggets of Strange Quark Matter (SQM), called
nuclearites when neutralized by captured electrons, and Q-balls. SQM
consists of aggregates of u, d and s quarks (in approximately 
equal proportions) will slighthy positive electric charge
\cite{nuclr}. It was suggested that SQM may be
the ground state of QCD. They should be stable for all baryon
numbers in the range between ordinary heavy nuclei and neutron stars
(A$\sim$ 10$^{57}$). 
They could have been produced in the early Universe or in violent
astrophysical processes. Nuclearite interaction with matter depend on their 
mass and size. In \cite{SLIM05/5} different mechanisms of energy loss
and propagation in relation to their detectability with the SLIM
apparatus are considered. In the absence of any
candidate, SLIM will be able to rule out some of the hypothesized
propagation mechanisms. Q-balls are super-symmetric coherent states of
squarks, sleptons and Higgs fields, predicted by minimal
super-symmetric generalizations of the Standard Model \cite{qballs}
they could have been produced in the early Universe. Charged Q-balls
should interact with matter in ways not too dissimilar from those of
nuclearites.
  
In the followings, after a short description of the apparatus, we
present the calibrations, the analysis procedures and preliminary results from
the SLIM experiment.

\section{Experimental}
The SLIM (Search for LIght magnetic Monopoles) experiment, based on 440
m$^2$ of Nuclear Track Detectors (NTDs), was deployed at the
Chacaltaya High Altitude 
Laboratory (Bolivia, 5260 m a.s.l.) since 2001 \cite{NTDsM}. Another
100 m$^2$ of NTDs were installed at Koksil (Pakistan, 4600 m a.s.l.) since
2003. The detector modules have been exposed under the roof of the
Chacaltaya Lab. at a height of 4 m above ground. The air
temperatures are recorded 3 times a day together with the minimum and
maximum values. From the observed ranges 
of temperatures we conclude that no
significant time variations occurred in the detector
response. The radon activity and the flux of cosmic ray neutrons were 
measured by us and by other authors \cite{neutron}.  

Extensive test studies were made in order to improve the etching
procedures of CR39 and Makrofol NTDs, improve the scanning and
analysis procedures and speed, and keep a good scan efficiency.  
"Strong" and "soft" etching conditions have been defined
\cite{NTDsM}. CR39 strong etching conditions (8N KOH + 1.25\%
Ethyl alcohol at 77$^\circ$ C for 30 hours) are used for the first
CR39 sheet in each module, in order to produce large tracks, easier to
detect during scanning. CR39 soft etching conditions (6N NaOH +
1\% Ethyl alcohol at 70$^\circ$ C for 40 hours) are applied to the other
CR39 layers in a module, if a candidate track is found in the first
layer. It allows more reliable measurements of the restricted energy
loss (REL) and of the direction of the incident particle. Makrofol
layers are etched in 6N KOH + Ethyl alcohol (20\% by volume), at
50$^\circ$ C. 
\begin{figure}[h]
 \centering
 {\centering\resizebox*{!}{7cm}{\includegraphics{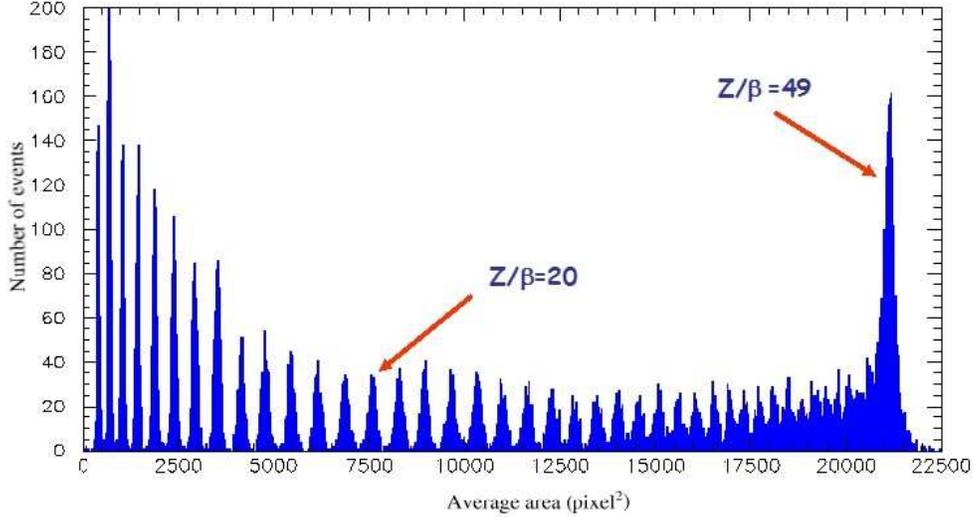}}\par}
\begin{quote}
 \caption{\label {fig1} Calibrations of CR39 nuclear track detectors
 with 158 A GeV In$^{49+}$ ions and their fragments (2 face measurements).} 
\end{quote}
 \end{figure}
  
The detectors have been calibrated using 158 A GeV $^{49+}$In (see
Fig. \ref{fig1}) and 30 A GeV $^{82+}$Pb beams at 
the CERN SPS. For soft etching conditions the threshold in CR39 is
at REL $\sim$ 50 MeV cm$^2$ g$^{-1}$; for strong etching the threshold
is at  REL $\sim$ 200 MeV cm$^2$ g$^{-1}$.  
Makrofol has a higher threshold (REL$ \sim$ 2.5 GeV
cm$^2$ g$^{-1}$). The CR39 allows the detection of IMMs  with two units
Dirac charge in the whole $\beta$-range of 4 10$^{-5}$ $<$ $\beta$ $<$
1. The Makrofol is useful for the detection of fast MMs, and nuclearites with
$\beta \sim$ 10$^{-3}$ can be detected by both CR39 and Makrofol. 	
\begin{figure}[h]
\begin{center}
\includegraphics[width=.73\textwidth]{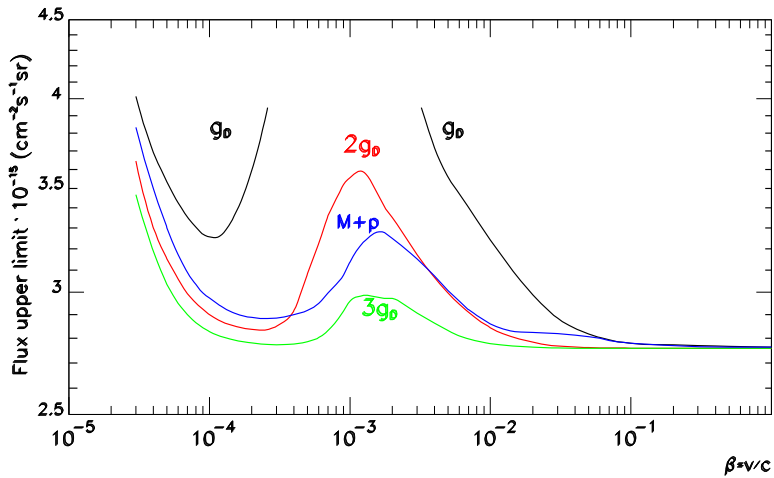}
\vspace{-0.6cm}
\begin{quote} 
\caption{\label {fig2} Present 90\% CL upper limits for a downgoing
 flux of IMMs with g = g$_{D}$, 2g$_{D}$, 3g$_{D}$ and for dyons
 (M+p , g = g$_{D}$) plotted vs $\beta$.}
\end{quote}
\end{center}
 \end{figure}

The analysis of a SLIM module starts by etching the top CR39
sheet using  strong conditions, reducing its thickness from 1.4 mm
to $\sim$ 0.6 mm. Since MMs, 
nuclearites and Q-balls should have a constant REL through the stack,
the signal looked for is a hole or a biconical track with the
two base-cone areas equal within the experimental uncertainties. The
sheets are scanned with a low magnificaton stereo
microscope. Possible candidates are further analysed with a high 
magnification microscope. The size of surface tracks is measured on
both sides of the sheet.  We require the two values to be equal within
3 times the standard deviation of their difference. A track is
defined as  a "candidate" if the REL and the incidence angles on the
front and back sides are equal to within 15\%.
To confirm the candidate track, the bottom CR39 layer is then
etched in soft conditions; an accurate scan under an
optical microscope with high magnification is performed  in a
region of about 0.5 mm around the expected candidate
position. If a two-fold coincidence is found the middle layer
of the CR39 (and in case of high Z candidate, the Makrofol layer) is
analyzed with soft conditions. No two-fold coincidence was found, that
is no MM, nuclearite or Q-ball candidate was detected. 

\section{Results and Conclusions}
We etched and analysed 236 m$^2$ of CR39, with an average exposure
time of more than 3.6 years. No candidate passed the search criteria: the 90\%
C.L. upper limits for a downgoing flux of fast ($\beta>$0.1) 
IMM's, nuclearites and Q-balls of any speed, all coming from above,
are at the level of $2.76\cdot 10^{-15}$ cm$^{-2}$ sr$^{-1}$ s$^{-1}$ (see
Fig. \ref{fig2}).
  
By the end of 2006 the 440 m$^2$ will be completely analyzed and the
experiment will reach a sensitivity of $\sim$ 10$^{-15}$ cm$^{-2}$
sr$^{-1}$ s$^{-1}$ for IMMs with $\beta \geq$ 10$^{-2}$; the same
sensitivity should be reached also for nuclearites and Q-balls with
galactic velocities. Moreover this search will benefit from the
analysis of further 100 m$^2$ of NTDs  installed at Koksil.

We acknowledge the collaboration of E. Bottazzi, L. Degli Esposti,
G. Grandi and C. Valieri of INFN Bologna and the
technical staff of the Chacaltaya Laboratory. We thank
INFN and ICTP for providing grants for non-italian citizens.

}
\end{document}